\documentstyle[aps,multicol,prl,epsfig]{revtex}
\begin{document}
\draft


\title{Force distributions near the jamming and glass transitions}
\author{Corey S. O'Hern$^{1,3}$, Stephen A. Langer$^2$, Andrea J. Liu$^1$,
and Sidney. R. Nagel$^3$}
\address{$^1$~Department of Chemistry and Biochemistry,
University of California at Los Angeles, \\ Los Angeles, CA  90095-1569}
\address{$^2$~Information Technology Laboratory, NIST,
Gaithersburg, MD 20899-8910}
\address{$^3$~James Franck Institute, The University of Chicago,
Chicago, IL 60637}
\date{\today}
\maketitle

\begin{abstract}
We calculate the distribution of interparticle normal forces $P(F)$
near the glass and jamming transitions in model supercooled liquids
and foams, respectively.  $P(F)$ develops a peak that appears near the
glass or jamming transitions, whose height increases with decreasing
temperature, decreasing shear stress and increasing packing density.
A similar shape of $P(F)$ was observed in experiments on static
granular packings.  We propose that the appearance of this peak
signals the development of a yield stress.  The sensitivity of the
peak to temperature, shear stress and density lends credence to the
recently proposed generalized jamming phase diagram.
\end{abstract}


\begin{multicols}{2}
\narrowtext

Granular materials can flow when shaken, but jam when the shaking
intensity is lowered\cite{granular}.  Similarly, foams and emulsions
can flow when sheared, but jam when shear stress is
lowered\cite{foam}.  These systems are athermal because thermal energy
is insufficient to change the packing of grains, bubbles, or droplets.
When the external driving force is too small to cause particle
rearrangements, these materials become amorphous solids and develop a
yield stress.  A supercooled liquid, on the other hand, is a {\it
thermal} system that turns, as temperature is lowered, into a
glass---an amorphous solid with a yield stress\cite{glass}.  Despite
significant differences between driven, athermal systems and
quiescent, thermal ones, it has been suggested that the process of
jamming---developing a yield stress in an amorphous state---may lead
to common behavior, and that these systems can be unified by a jamming
phase diagram\cite{liunagel}.  This implies that there should be
similarities in these different systems as they approach jamming
or glass transitions.  We test this speculation by measuring the
distribution $P(F)$ of interparticle normal forces $F$, in model
supercooled liquids and foams.  We find that for glasses, $P(F)$ is
quantitatively similar to experimental results on granular
materials\cite{mueth}.

When granular materials jam, the distribution of stresses is known to
be inhomogeneous\cite{dantu,cliu}.  As proposed in Ref.~\cite{cliu},
we quantify this effect by measuring $P(F)$.  Our aim is to determine
which feature in $P(F)$ is associated with development of a yield
stress.  Experiments\cite{mueth,lovoll} and simulations
\cite{simulations,makse} on static granular packings find that $P(F)$
has a plateau or small peak at small $F$ and decays exponentially at
large $F$.  We argue that the development of a peak is the signature
of jamming.

For supercooled liquids, equilibrium statistical mechanics gives
insight into the shape of $P(F)$.  Since forces depend only on
particle separations, $P(F) dF = G(r) dr$, where $G(r) dr$ is the
probability of finding a particle between $r$ and $r+dr$ given a
particle at the origin.  Thus, $G(r)=\rho/(N-1) S_{D} r^{D-1} g(r)$,
where $N$ is the number of particles, $\rho$ is the number density,
$g(r)$ is the pair distribution function, and $S_{D} r^{D-1}$ is the
surface area of a $D$-dimensional sphere of radius $r$.  Although it
is well known that $g(r)$ does not change significantly as the
temperature is varied through the glass transition $T_{g}$, we show
below that $P(F)$ is quite sensitive and actually develops a peak near
$T_{g}$.  Physically, forces (or stresses\cite{alexander,perera}) are
crucial for understanding the slowing down of stress relaxation near
the glass transition, or the development of a yield stress.  It is
therefore not surprising that $P(F)$, which is a particular weighting
of $g(r)$, is much more sensitive to the glass transition than $g(r)$
itself.

In a jammed system like a granular material, an analytic
expression for $g(r)$ is not known and $P(F)$ must be measured
directly.  However, in an {\it equilibrium} system at temperature $T$,
the large-force behavior of $P(F)$ can be obtained from the
small-separation behavior of $g(r)$: $g(r) = y(r) \exp[-V(r)/k_bT]$,
where $V(r)$ is the pair potential and $y(r)$ depends relatively weakly on
$r$ at small
$r$\cite{widom}.
This leads to:
\begin{equation}
P[F(r)] \sim y(r) r^{D-1} {dr \over dF}
\exp\left[-{V(r)/k_b T}\right].
\label{highforce}
\end{equation}

{}From our simulations, we compute the force distributions
in systems that are out of equilibrium, such as glasses and sheared
foams, as well as systems in equilibrium, such as supercooled liquids.
We find that $P(F)$ for supercooled liquids (with
sufficiently strong repulsive potentials) decays approximately
exponentially at large forces, as predicted by Eq.~\ref{highforce}.
Because this is true at all
temperatures, even those far above $T_{g}$, the exponential tail is
not necessarily a signature of an amorphous solid.

We perform constant-temperature molecular dynamics simulations on
binary mixtures in $2$D, using the Gaussian constraint thermostat and
leapfrog Verlet algorithm\cite{allen}.  The masses $m$ of the
particles are the same, but the ratio of particle diameters,
$\sigma_2/\sigma_1 =1.4$, ensures that the system does not
crystallize\cite{perera}.  We confine $N=1024$ particles (512 of each
variety) to a square box and use periodic boundary conditions.  For
each simulation, we choose one of the following interparticle pair
potentials:
\begin{eqnarray}
\label{potentialdef}
V^{SC}_{ab}(r) & \equiv & \epsilon \left({\sigma_{ab}/r}\right)^{12} \\
%
V^{LJ}_{ab}(r) & \equiv &  4\epsilon\left[ \left({\sigma_{ab}/r}
\right)^{12} -
\left({\sigma_{ab}/r}\right)^6 \right] \nonumber \\
%
V^{LJR12}_{ab}(r) & \equiv & 4\epsilon\left[
({\sigma_{ab}/r})^{12} - ({\sigma_{ab}/r})^6 \right] +
\epsilon;~~ {r/\sigma_{ab}} \le 2^{1/6} \nonumber \\
V^{LJR24}_{ab}(r) & \equiv &  {2^{8/3}\epsilon\over3}
\left[ ({\sigma_{ab}/r})^{24} - ({\sigma_{ab}/r})^6 \right]
+ \epsilon;\nonumber\\
&&\qquad\qquad\qquad\qquad\qquad\qquad {r/\sigma_{ab}} \le 2^{1/9} \nonumber
\end{eqnarray}
where $\sigma_{ab} = (\sigma_a +\sigma_b)/2$ for $a,b=1,2$.  The
potentials $LJR12$ and $LJR24$ are zero above the specified cutoffs.
(The potentials $SC$ and $LJ$ are truncated at large $r$,
$r/\sigma_{ab}=4.5$.)   Below, we measure time, force, temperature, and density
in units of $\sigma_1(m/\epsilon)^{1/2}$, $\epsilon/\sigma_1$,
$\epsilon/k_b$, and $\sigma_1^{-2}$ respectively.  The simulations on
purely repulsive potentials $SC$, $LJR12$,
and $LJR24$ simulations were carried out at constant density
$\rho=0.747$; the simulations on $LJ$, which includes an attraction,
were carried out at zero average pressure.

The hallmark of the glass transition is the extreme slowing down of
the dynamics as temperature is lowered toward the glass transition.
The pair
potentials in Eq.~\ref{potentialdef} all give rise to glass
transitions as temperature is lowered\cite{perera}.  We determine
$T_g$ by measuring the self-part of the intermediate scattering
function $F_{2}(k_p,t)$ for the large particles at a wavevector $k_p$
corresponding to the first peak of the static structure
factor\cite{perera,kob}.  For high temperatures, the liquid
equilibrates quickly and $F_{2}(k_p,t)$ decays exponentially to zero.
The relaxation time $\tau_r$ is defined as the time at which
$F_{2}(k_p,t)$ decays to $1/e$; this is a measure of the
$\alpha$-relaxation time\cite{perera,kob}.  Since $\tau_{r}$ increases so
rapidly near the glass transition,
simulations can only reach equilibrium for
temperatures $T > T_g$, where $T_g$ is determined by when $\tau_{r}$
exceeds a predetermined, large value, which we take to be $\tau_r > 1000$.
For our
parameter choices, the glass transition temperatures are
$T_{g}^{SC}=0.38$, $T_{g}^{LJR12}=1.1$, $T_{g}^{LJR24}=3.0$ and
$T_{g}^{LJ}=0.17$.

\begin{figure}
\epsfxsize=2.6in \epsfbox{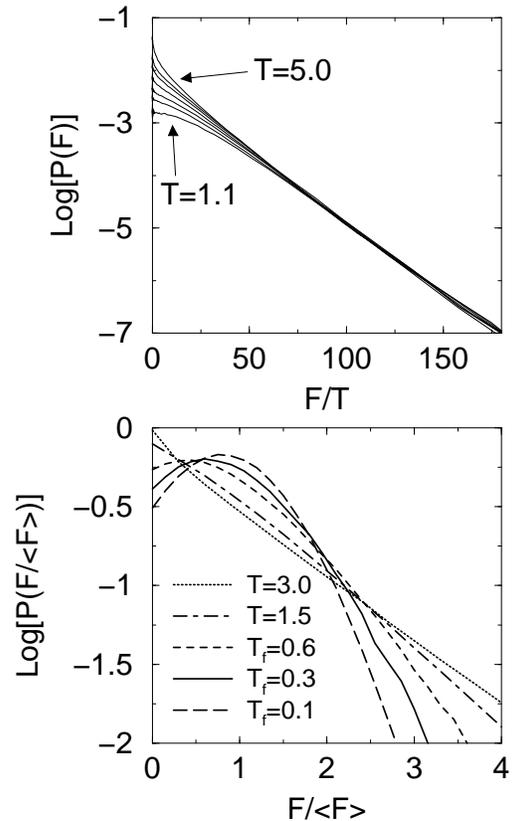}
\caption{Top: $P(F)$ for all interparticle force pairs versus $F/T$
for LJR$12$ obtained for seven temperatures above $T_g$ with $T$
decreasing from top to bottom.  Bottom: $P(F/\langle F\rangle)$ versus
$F/\langle F\rangle$ for LJR$12$ for two temperatures above and three
below $T_g$.}
\label{pf}
\end{figure}

For $T>T_g$ we measure $P(F)$ for all interparticle force pairs from
at least $250$ configurations after equilibrating each configuration
for $10-100\tau_r$.  The top frame of Fig.~\ref{pf} shows $P(F)$
plotted versus $F/T$ for $LJR12$ for seven temperatures above $T_g$ with
temperature decreasing from top to bottom.  At high temperatures, we
see in the top frame of Fig.~\ref{pf} that $P(F)$ increases with
decreasing $F$ over the entire range of $F$.  However, as temperature
is lowered towards $T_g$, a plateau in $P(F)$ forms at forces below
the average $\langle F \rangle$.  By shifting each curve vertically,
we have obtained collapse of the high-force data for all of these {\it
equilibrium} systems.  {}From Eq.~\ref{highforce}, we expect that the
large-force tail should scale asymptotically as $\exp(-B
F^{12/13}/T)$, where $B$ is a constant and the power $12/13$ derives
from the $1/r^{12}$ repulsion.  Thus, for particles with harder cores
(steeper repulsions), the tail becomes closer to an exponential in
$F$, as seen in experiments on granular materials\cite{mueth}.  This
explanation for the exponential tail is different from that of the
q-model\cite{cliu} and its generalizations\cite{clement} based on
stochastic force propagation.  Previous LJ simulations along the
liquid-vapor coexistence line \cite{powles} showed that the Cartesian
components of the force also have an exponential distribution.  Our
results are related to theirs: for high forces, the total force on a
particle, which is the vector sum of the normal forces, will be
dominated by the largest normal force.  This is why the distribution
of Cartesian components is also exponential.

We also study $P(F)$ {\it out} of equilibrium by performing thermal
quenches from $T_i > T_g$ to $T_f < T_g$.  The results discussed below
are relatively insensitive to changes in $T_i$ or quench rate.  In the
bottom frame of Fig.~\ref{pf}, we show the long-time behavior of
$P(F)$ for $LJR12$ following a quench below $T_{g} \approx 1.1$ to
$T_f=0.6$, $0.3$, and $0.1$.  For comparison, two equilibrium
distributions at $T=1.5$ and $3.0$ are also shown.  We scaled the
abscissa by $\langle F\rangle$ (which increases with $T$).  There are
two significant features in $P(F)$ for glasses.  First, the slope of
the exponential tail increases as $T$ is lowered.  The temperature
corresponding to the tail $T_{\rm tail}$, however, is not the final
temperature $T_{f}$, but rather satisfies $T_{f}<T_{\rm tail}<T_{g}$.
Thus, a fraction of the large thermal forces cannot relax in the
glassy state.  The second significant feature of $P(F)$ for glasses is
the formation of a peak near $\langle F\rangle$, as shown in the
bottom frame of Fig.~\ref{pf}.  Thus, in contrast to $g(r)$, there is
a significant change in $P(F)$ below $T_g$.  The behavior of $P(F)$
when quenched below $T_g$ is qualitatively the same for all potentials
and densities studied, showing that the peak signals the glass
transition in a system with attractive interactions and no applied
pressure as well as systems with purely repulsive interactions under
pressure.

\begin{figure}
\centerline{\epsfxsize=3.0in \epsfbox{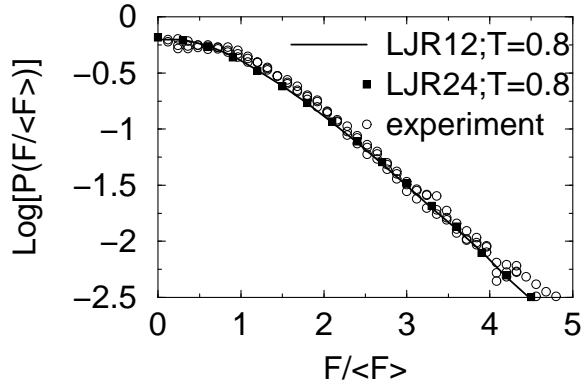}}
\caption{$P(F/\langle F\rangle)$ versus $F/\langle F\rangle$ for both
LJR$12$ and LJR$24$ after a quench to $T_f=0.8$.  Data from
experiments on static granular packings from \protect\cite{mueth} are
also shown.  Note that all three sets of data have a plateau at small
$F$ and decay exponentially at large $F$.}
\label{experiment}
\end{figure}

The potentials $LJR12$ and $LJR24$ in Eq.~\ref{potentialdef} are most
similar to granular materials since they produce purely repulsive
forces that vanish at small separation.  We compare $P(F/\langle F
\rangle)$ in the
glassy state for $LJR12$ and $LJR24$ to $P(F/\langle F \rangle)$ for 
static granular
packings in Fig.~\ref{experiment}.  For both $LJR12$ and $LJR24$, we
have quenched to $T_f=0.8<T_g$.  Remarkably, the force distributions,
when scaled by the average force $\langle F\rangle$, are nearly
identical for $LJR12$, $LJR24$, and experiments on static granular
packings\cite{mueth} over the entire range of forces.  This implies
that for sufficiently hard repulsive potentials, the shape of the
distribution is not sensitive to the shape of the potential.  In the
limit of hard spheres, where the power of the repulsive term in the
potential diverges, we expect similar behavior.  In systems with
softer potentials, such as Hertzian or harmonic repulsive springs, we
also find the same shape of $P(F)$ as in Fig.~\ref{experiment}
at very low temperatures near the
close-packing density\cite{ohern}.  These results suggest that
the slight peak or plateau at small forces and exponential tail at
large forces are a generic feature of $P(F)$ in athermal, experimental
systems near the onset of jamming.

\begin{figure}
\centerline{\epsfxsize=3.0in \epsfbox{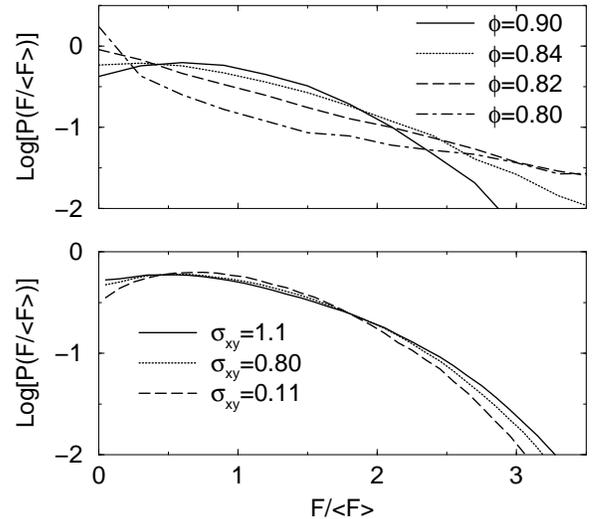}}
\caption{Top: $P(F/\langle F\rangle)$ versus $F/\langle F\rangle$ for foams
with $\sigma_{xy}=0$ for several $\phi$ near random close-packing.
Bottom: $P(F/\langle F\rangle)$ for foams with $\phi=0.9 >\phi_0$ and
$\sigma_{xy}$ lowered towards $\sigma_y$.  }
\label{pffoam}
\end{figure}

Is a peak or plateau in $P(F)$ also observed in other jammed systems?
To answer this, we have studied model two-dimensional
foams\cite{durian,liulanger}, where bubbles are treated as circles
that can overlap and interact via two types of pairwise interactions.
The first is a harmonic repulsion that is nonzero when the distance
between centers of two bubbles is less than the sum of their radii.
The other is a simple dynamical friction proportional to the relative
velocities of two neighboring bubbles.  In foam, thermal motion of
bubbles is negligible.  We simulate a $400$-bubble system at constant
area with periodic boundary conditions in the $x$-direction and fixed
boundaries in the $y$-direction.  Bubble radii $R_i$ are chosen from a
flat distribution with $0.2< R_i/\langle R\rangle< 1.8$.

At packing fractions $\phi$ above random close packing ($\phi_0
\approx 0.84$), quiescent foam is an amorphous solid with a yield
stress $\sigma_{y}$.  However, when shear stress
$\sigma_{xy}>\sigma_{y}$ is applied, the foam flows.  There are
therefore two ways to approach the amorphous solid.  We can either
increase $\phi$ towards $\phi_{0}$ at $\sigma_{xy}=0$ (route $1$), or
we can decrease $\sigma_{xy}$ towards $\sigma_y$ at fixed
$\phi>\phi_{0}$ (route 2).  In Fig.~\ref{pffoam}, we show $P(F/\langle
F\rangle)$ (only including harmonic elastic forces) along these two
routes.  The distributions along route $1$ in the top frame were
measured after quenching $50$ configurations from $\phi_i \ll
\phi_{0}$ to $\phi$ by increasing each particle radius.  When
$\phi<\phi_{0}$, $P(F/\langle F \rangle)$ increases monotonically as
$F/\langle F\rangle$ decreases.  As $\phi$ increases above $\phi_{0}$,
a local maximum forms near $\langle F\rangle$.  A similar trend is
found along route $2$.  To obtain these distributions, we averaged
over at least $500$ configurations with each brought to steady state
for a strain of $\approx 10$.  In all cases shown, $\sigma_{xy}$
exceeds $\sigma_y$, so the systems are flowing.  We find that at large
$\sigma_{xy}$, $P(F/\langle F \rangle)$ is nearly constant at small
$F$.  When $\sigma_{xy}$ is lowered towards $\sigma_y \approx 0.10$, a
peak in $P(F/\langle F\rangle)$ forms near $F/\langle F\rangle \sim
1$.  Similar behavior is observed in $P(F/\langle F\rangle)$ as a
function of $\phi$ in experiments on sheared deformable disks
\cite{howell} and as a function of confining stress in simulations of
deformable spheres\cite{makse}.

In this Letter, we have shown a connection between development of a
yield stress, either by a glass transition or conventional jamming
transition, and the appearance of a peak in $P(F)$.  We have
established that four different model supercooled liquids develop
first a plateau and then a peak in $P(F)$ as temperature is lowered
below the glass transition.  We have also found (but not shown here)
that the $LJR12$ liquid displays identical results for $P(F)$ as shear
stress is lowered from the flowing state or as density is raised from
the liquid state at fixed temperature\cite{ohern}.  We also find that
the appearance of a peak in $P(F)$ coincides with onset of
crystallization in monodisperse liquids, further supporting the
connection between the peak and the yield stress\cite{ohern}.  The
athermal foam likewise develops a peak in $P(F)$ as it approaches
jamming along two different routes.  Static granular packings exhibit
a plateau or small peak in $P(F)$ as well.  Thus, a peak in $P(F)$
appears as a wide variety of systems jam along each of the axes of the
jamming phase diagram\cite{liunagel}.  This suggests that jamming
leads to common behavior and that the glass transition may resemble
more conventional jamming transitions.

This still leaves the question of why formation of a peak or plateau
in $P(F)$ appears to signal the development of a yield stress.  The
presence of the peak or plateau implies that there are a large number
of forces near the average value.  This is consistent with the
existence of force chains, since each particle within a force chain
must have roughly balanced forces on either side.  We speculate that
systems jam when there are enough particles in a force chain network
to support the stress over the time scale of the measurement.  
Forces at the peak of $P(F)$ are among the slowest to relax: these
forces correspond to separations near the first peak of $g(r)$, which
stem from wavevectors near the first peak in the static structure
factor, which are among the most slowly relaxing modes\cite{pgdg}.
This implies that force chains observed in granular packings may also
be important to the glass transition.  The fact that force chains do
not couple strongly to density fluctuations may explain why they have
not been observed directly.  However, large kinetic heterogeneities
that appear near $T_{g}$\cite{ediger} may be linked to the formation
of force chains.  This interpretation suggests that force chains
may provide the key to the elusive order parameter for the glass
transition.

We thank Susan Coppersmith, Heinrich Jaeger, Robert Leheny, Daniel
Mueth, Thomas Witten, Walter Kob, and Gilles Tarjus for instructive
discussions.  Support from NSF Grant Nos. DMR-9722646 (CSO,SRN),
CHE-9624090 (CSO,AJL), and PHY-9407194 (SAL,AJL,SRN) is gratefully
acknowledged.

\end{multicols}

\begin{references}

\bibitem{granular}
H. M. Jaeger, S. R. Nagel, and R. P. Behringer, Rev. Mod. Physics
{\bf 68}, 1259 (1996).

\bibitem{foam} D. J. Durian and D. A. Weitz,
"Foams," in Kirk-Othmer Encyclopedia of Chemical Technology, 4 ed.,
ed. J.I. Kroschwitz (Wiley, New York, 1994), Vol. 11, p. 783.

\bibitem{glass}
M. D. Ediger, C. A. Angell, and S. R. Nagel, J. Phys. Chem. {\bf 100},
13200 (1996).

\bibitem{liunagel}
A. J. Liu and S. R. Nagel, Nature {\bf 396} (N6706), 21 (1998).

\bibitem{mueth}
D. M. Mueth, H. M. Jaeger, and S. R. Nagel, Phys. Rev. E {\bf 57},
3164 (1998); D. L. Blair, N. W. Mueggenburg, A. H. Marshall, H. M. Jaeger,
and S. R. Nagel, (unpublished).

\bibitem{dantu}
P. Dantu, G\'{e}otechnique {\bf 18}, 50 (1968).

\bibitem{cliu}
C.-h. Liu, S. R. Nagel, D. A. Schecter, S. N. Coppersmith, S. Majumdar,
O. Narayan, and T. A. Witten, Science {\bf 269}, 513 (1995);
S. N. Coppersmith, C.-h. Liu, S. Majumdar, O. Narayan, and T. A. Witten,
Phys. Rev. E {\bf 53}, 4673 (1996).

\bibitem{lovoll}
G. L$\o$voll, K. J. M{\aa}l$\o$y, and E. G. Fekk$\o$y,
Phys. Rev. E {\bf 60}, 5872 (1999).

\bibitem{simulations} F. Radjai, M. Jean, J.-J. Moreau, and S. Roux,
Phys. Rev. Lett. {\bf 77}, 274 (1996); S. Luding, Phys. Rev. E {\bf
55}, 4720 (1997); A. V. Tkachenko and T. A. Witten, Phys. Rev. E {\bf
60}, 687 (1999).

\bibitem{makse} H. A. Makse, D. L. Johnson and L. M. Schwartz,
Phys. Rev. Lett. {\bf 84}, 4160 (2000); C. Thornton, KONA Powder
Part. {\bf 15}, 81 (1997).

\bibitem{alexander} S. Alexander, Phys. Rep. {\bf 296}, 65 (1998); T.
Kustanovich, S. Alexander, Z. Olami, Physica A {\bf 266}, 434 (1999); T.
Kustanovich and Z. Olami, Phys. Rev. B {\bf 61}, 4813 (2000).

\bibitem{widom}
B. Widom, J. Phys. Chem. {\bf 86}, 869 (1982); L. L. Lee, D. Ghonasgi,
E. Lomba, J. Chem. Phys. {\bf 104}, 8058 (1996).

\bibitem{allen}
M. P. Allen and D. J. Tildesley, {\it Computer Simulations of Liquids},
(Oxford University Press,Oxford,1987).

\bibitem{perera}
D. N. Perera and P. Harrowell, Phys. Rev. E {\bf 59}, 5721 (1999).

\bibitem{kob}
W. Kob and H. C. Anderson, Phys. Rev. E {\bf 52}, 4134 (1995).

\bibitem{clement} C. Eloy and E. Clement, J. Phys. I (France) {\bf
7}, 1541 (1997); J. E. S. Socolar, Phys. Rev. E {\bf 57}, 3204
(1998); P. Claudin, J. -P. Bouchaud, M. E. Cates, J. -P. Wittmer, Phys.
Rev. E {\bf 57}, 4441 (1998); M. L. Nguyen and S. N. Coppersmith,
Phys. Rev. E {\bf 59}, 5870 (1999).

\bibitem{powles}
J. G. Powles and R. F. Fowler, Mol. Phys. {\bf 62}, 1079 (1987).

\bibitem{durian}
D. J. Durian, Phys. Rev. E {\bf 55}, 1739 (1997).

\bibitem{liulanger}
S. A. Langer and A. J. Liu, Europhys. Lett. {\bf 49}, 68 (2000);
S. Tewari, D. Schiemann, D. J. Durian, C. M. Knobler, S. A. Langer, and
A. J. Liu, Phys. Rev. E {\bf 60}, 4385 (1999).

\bibitem{howell} D. Howell and R. P. Behringer, Phys. Rev. Lett. {\bf 82},
5241 (1999).

\bibitem{ohern}
C. S. O'Hern, S. A. Langer, A. J. Liu, and S. R. Nagel (unpublished).

\bibitem{pgdg}
P.-G. de Gennes, Physica {\bf 25}, 825 (1959).  For applications
to simulations of glass-forming liquids, see \protect\cite{kob}.

\bibitem{ediger} M. D. Ediger, Ann. Rev. Phys. Chem. (in press).

\end{references}
\end{document}